\documentstyle[pre,aps]{revtex}

\newcommand{\be}{\begin{equation}}
\newcommand{\ee}{\end{equation}}
\newcommand{\bB}{{\bf B}}
\newcommand{\bS}{{\bf S}}
\newcommand{\gm}{\gamma}
\newcommand{\bt}{\beta}
\newcommand{\al}{\alpha}

\newcommand{\om}{\omega}
\newcommand{\sgm}{\sigma}

\newcommand{\Dlt}{\Delta}
\newcommand{\Gm}{\Gamma}

\begin{document}

\draft

\title{Processing Information by Punctuated Spin Superradiance}

\author{V.I. Yukalov$^{1,2}$ and E.P. Yukalova$^{1,3}$} 

\address{$^1$Research Center for Optics and Photonics \\
Instituto de Fisica de S\~ao Carlos, Universidade de S\~ao Paulo \\
Caixa Postal 369, S\~ao Carlos, S\~ao Paulo 13560-970, Brazil}

\address{$^2$Bogolubov Laboratory of Theoretical Physics \\
Joint Institute for Nuclear Research, Dubna 141980, Russia}

\address{$^3$Department of Computational Physics \\
Laboratory of Informational Technologies \\
Joint Institute for Nuclear Research, Dubna 141980, Russia}

\maketitle

\vskip 2cm

\begin{abstract}

The possibility of realizing the regime of punctuated spin 
superradiance is advanced. In this regime, the number of superradiant 
pulses and the temporal intervals between them can be regulated. This 
makes it feasible to compose a kind of the Morse Code alphabet and, 
hence, to develop a technique of processing information.
\end{abstract}

\vskip 2cm


Spin systems can exhibit a phenomenon that is analogous to atomic
superradiance [1,2], because of which it is called spin superradiance.
To realize this phenomenon, spin systems, similarly to atomic ones,
are to be prepared in an inverted state. This is achieved by placing 
a polarized spin sample in an external magnetic field directed opposite
to spin polarization. Contrary to atomic systems, coherent spin motion
develops not owing to direct spin correlations but due to the 
interaction of spins with a resonator feedback field, for which purpose
the spin sample has to be coupled with a resonant electric circuit,
whose natural frequency is tuned to the Zeeman frequency of spins [3].
More details on similarities and differences between atomic 
superradiance and spin superradiance can be found in the review [4].
Spin superradiance is the process of {\it coherent spontaneous emission}
by moving spins. As in the case of atomic systems, one may distinguish
two main types of this phenomenon, transient superradiance and pulsing
superradiance. {\it Transient spin superradiance} occurs when the spin
sample is prepared in the inverted state, after which no following 
pumping is involved. In this case, a single superradiant burst arises,
peaked at the delay time. {\it Pulsing spin superradiance}
is radically different from the transient regime by the occurrence of
a series of superradiant pulses, for which the spin sample is to be 
subject to a permanent pumping supporting the inverted spin polarization.
Both regimes of spin superradiance, transient [5--8] as well as pulsing
[9--11] were observed in experiments with different materials containing
nuclear spins. A microscopic theory of these phenomena, based on
realistic spin Hamiltonians, was developed [12--15], being in good
agreement with experiment and with computer modelling [16]. It is worth
stressing that only by invoking microscopic Hamiltonians it has become 
possible to give an accurate description of purely self-organized
regimes which cannot be treated by the phenomenological Bloch equations 
[12--14].

In the present paper, we advance the possibility of realizing the 
third type of spin superradiance, which we call {\it punctuated spin 
superradiance}, and which is principally different from the transient
and pulsing types. In this regime, unlike the transient case, not a
single but many superradiant bursts can be produced. In distinction to
the pulsing regime, where the number of pulses and the temporal distance
between them are prescribed by a given setup and cannot be varied, in
the process of punctuated superradiance both the number of superradiant
bursts as well as time intervals between each pair of them can be 
regulated. The term {\it punctuation} here means this feasibility of
changing interpulse intervals and of organizing various groups of
superradiant bursts. In that way, a code, like the Morse alphabet, 
can be composed, which may be employed in processing information.

The consideration below will be based on the Hamiltonian typical for 
spin systems employed in magnetic resonance [17,18]. The Hamiltonian reads
\be
\label{1}
\hat H = \sum_i \hat H_i  + \frac{1}{2}\; \sum_{i\neq j} \hat H_{ij}\; ,
\ee
where $\hat H_i$ corresponds to individual  spins, while 
$\hat H_{ij}$, to spin interactions, with the indices $i,j=1,2\ldots,N$ 
enumerating spins. The individual terms are given by the Zeeman energy
$\hat H_i = -\mu_0\bB\cdot\bS_i$, where $\mu_0\equiv\hbar\gm_S$, with
$\gm_S$ being the gyromagnetic ratio of spin $S$, represented by the 
spin operator $\bS_i$ and $\bB$ is the total magnetic field acting on 
each spin. The spin interactions are described by the dipolar terms
$\hat H_{ij} =\sum_{\al\bt} C_{ij}^{\al\bt} S_i^\al S_j^\bt$, 
with the dipolar tensor $C_{ij}^{\al\bt}$. The total magnetic field 
$\bB=B_0{\bf e}_z+H{\bf e}_x$ consists of a constant longitudinal
field $B_0$ and a transverse field $H$ formed by the resonant electric 
circuit coupled to the spin sample. The resonator field $H=4\pi nj/cl$ 
is created by the electric current $j$ circulating over a coil of $n$
turns and length $l$. The current $j$ is determined by the Kirchhoff
equation. The electric circuit is characterized by resistance $R$, 
inductance $L$, and capacity $C$. With the notation for the circuit
natural frequency $\om\equiv 1/\sqrt{LC}$ and circuit damping 
$\gm\equiv R/2L$, the Kirchhoff equation can be presented as 
\be
\label{2}
\frac{dH}{dt} + 2\gm H + \om^2 \int_0^t H(t')dt' = - 4\pi\eta \;
\frac{dM_x}{dt} 
\ee
for the resonator magnetic field $H$, where $\eta$ is a filling 
factor and $M_x=(\mu_0/V)\sum_i<S_i^x>$ is the $x$-component of the 
magnetization density of a sample with volume $V$. Since the resonator 
field $H$, acting on spins, is itself due to the transverse spin motion,
this field $H$ is called the feedback field.

To derive evolution equations, we follow the scale separation approach, 
described in full detail in Refs. [12--15]. For this purpose, we write 
the Heisenberg equations of motion for the lowering, $S_i^-$, raising,
$S_i^+$, and polarization, $S_i^z$, operators. In these equations, it is 
possible to separate the combinations describing fast fluctuating local 
fields. Employing the method of random local fields [17--19], the latter
are modelled by stochastic Gaussian variables, with zero mean and the 
width $2\gm_3$, where $\gm_3$ is the width of inhomogeneous dynamic 
broadening. Then the Heinsenberg equations are averaged over spin 
degrees of freedom, not touching the stochastic variables. Denoting the 
averaging over spins by single angle brackets $<\ldots>$, we define the 
transition function $x$, coherence intensity $y$, and spin polarization 
$z$, respectively,
\be
\label{3}
x\equiv \frac{1}{S}\; < S_i^-> \; , \qquad
y \equiv \frac{1}{S^2}\; < S_i^+><S_i^-> \; , \qquad
z\equiv \frac{1}{S}\; <S_i^z> \; .
\ee
The wavelength of spin radiation is usually much larger than 
interparticle distance, because of which the uniform approximation 
for Eqs. (3) may be employed.

We direct the external magnetic field $B_0$ so that $\mu_0B_0<0$ and 
define the Zeeman frequency $\om_0 \equiv|\mu_0 B_0|/\hbar$.
Also, introduce the notation $f \equiv -(i/\hbar)\mu_0 H + \xi$
for an effective force acting on spins. Finally, the evolution equations 
for the functions (3) can be cast [12--15] to the form
$$
\frac{dx}{dt} = - i(\om_0 +\xi_0 -i\gm_2) x + fz \; , \qquad
\frac{dy}{dt} = - 2\gm_2 y\left ( x^* f + f^* x\right ) z \; ,
$$
\be
\label{4}
\frac{dz}{dt} = -\; \frac{1}{2}\left ( x^* f + f^* x\right ) -
\gm_1(z -\sgm) \; ,
\ee
where $\gm_1$ and $\gm_2$ are the longitudinal and transverse widths, 
respectively, and $\sgm$ is an equilibrium polarization of a spin. When 
there is no external stationary pumping, $\sgm=-1$. Equations (4) are
stochastic differential equations, since they contain random fields.

To solve Eqs. (4), we invoke a generalization [12--14] of the averaging
technique [20] to the case of stochastic differential equations. This
becomes possible owing to the existence of several small parameters:
$\gm_0/\om_0\ll 1$, $\gm_1/\om_0\ll 1$, $\gm_2/\om_0\ll 1$, and
$\gm_3/\om_0\ll 1$, where $\gm_0\equiv \pi\eta\rho\mu_0^2 S/\hbar$ is 
the natural width and $\rho\equiv N/V$ is the density of spins. In 
addition, the resonant circuit, coupled to the spin sample, is assumed 
to be of good quality and tuned close to the Zeeman frequency $\om_0$, 
so that $\gm/\om\ll 1$, $|\Dlt|/\om\ll 1$, with $\Dlt\equiv\om -\om_0$.

First of all, using the occurrence of the small parameters, we can 
obtain an iterative solution of the Kirchhoff equation (2). For this
purpose, invoking the method of Laplace transforms, we present Eq. (2)
in the integral form
\be
\label{5}
H = -4\pi\eta \int_0^t G(t-t')\dot{M}_x(t') \; dt'\; , \qquad
G(t) =\left ( \cos\om't - \; \frac{\gm}{\om'}\; \sin\om't\right )
e^{-\gm t} \; , 
\ee
where the overdot means time differentiation and $\om'\equiv \sqrt{\om^2 -\gm^2}$. 
Since $M_x$ is expressed through $x$, its time derivative $\dot{M}_x$
is directly related to the first of Eqs. (4). Using this, we find the
solution of Eq. (5), in the first order with respect to small parameters, 
as $\mu_0H/\hbar=i\al(x-x^*)$, where the function
\be
\label{6}
\al = g\gm_2\left ( 1  - e^{-\gm t}\right ) \; , \qquad
g\equiv \frac{\gm\gm_0\om}{\gm_2(\gm^2+\Dlt^2)}
\ee
describes the intensity of coupling between the spin sample and the 
resonant circuit. Let us stress that the spin-resonator coupling (6)
depends on time, taking into account retardation effects.

From Eqs. (4), in the presence of the small parameters, it follows that 
the variables $y$ and $z$ are temporal quasi-invariants with respect to 
$x$. Then, we solve the first of Eqs. (4), with these quasi-invariants 
fixed, substitute the found solution $x$ into the second and third of 
Eqs. (4), and average the right-hand sides of these equations over time 
and over the stochastic fields. As a result, we obtain the guiding 
center equations
\be
\label{7}
\frac{dy}{dt} = - 2(\gm_2 -\al z) y + 2\gm_3 z^2 \; , \qquad
\frac{dz}{dt} = - \al y - \gm_3 z - \gm_1(z-\sgm) \; .
\ee

In the dynamics of solutions to Eqs. (7), one may distinguish two 
stages, quantum and coherent. At the {\it quantum stage}, when 
$\gm t\ll 1$, the coupling function is close to zero and no noticeable
coherence in the motion of transverse spins is yet developed. The
dynamics is governed by quantum spin interactions. At this stage, when
$\al\approx 0$, equations (7) are linear, and their solution is easy.
With increasing time, the spin-resonator coupling (6) grows, and
coherent effects, caused by the resonator feedback field, gradually
come into play. The crossover time between the quantum and coherent
regimes can be defined as the time $t_c$, at which the first term in
the first of Eqs. (7) changes its sign. This is because the quantity 
$\Gm\equiv\gm_2-\al z$ plays the role of an effective attenuation.
When the latter is positive, transverse coherence decays, while 
a negative attenuation implies the generation of coherence. Hence,
the moment of time, when $\Gm(t_c)$ changes its sign, separates 
qualitatively different regimes of spin motion. The {\it crossover time}
$t_c$, defined by the equality $\al(t_c)z(t_c)=\gm_2$, is
$t_c =\tau\ln[gz_0/(gz_0 -1)]$, with $\gm\tau =1$. The solutions $y$ and 
$z$ at this boundary of the quantum stage are $y(t_c)\simeq y_0 + 2\gm_3
t_c z_0^2$ and $z(t_c) \simeq z_0 +\gm_1 t_c\sgm$, where $y_0=y(0)$ and 
$z_0=z(0)$. The {\it coherent stage} of motion comes after the crossover
time $t_c$, when the spin-resonator coupling $\al$ fastly grows to 
$g\gm_2$. Since $\gm_3\leq\gm_2$, and if $g\gg 1$, then $g\gm_2\gg\gm_3$,
and the term with $\gm_3$ can be omitted. In the transient regime, when
$t\ll T_1\equiv\gm_1^{-1}$, the term containing $\gm_1$ can also be 
neglected. Then Eqs. (7) possess the exact solution
\be
\label{8}
y=\left( \frac{\gm_p}{g\gm_2}\right )^2{\rm sech}^2\left (
\frac{t-t_0}{\tau_p}\right ) \; , \qquad
z = -\; \frac{\gm_p}{g\gm_2}\; {\rm tanh}\left (
\frac{t-t_0}{\tau_p} \right ) + \frac{1}{g} \; ,
\ee
in which the pulse time $\tau_p$ and the pulse width $\gm_p$, 
with $\gm_p\tau_p\equiv 1$, are defined by the expressions
$\gm_p^2=\gm_g^2+(g\gm_2)^2(y_0+2\gm_3 t_c z_0^2)$, 
$\gm_g\equiv\gm_2(1-g z_0)$, and the delay time is
\be
\label{9}
t_0 = t_c + \frac{\tau_p}{2} \ln\left | 
\frac{\gm_p - \gm_g}{\gm_p + \gm_g} \right | \; .
\ee
Solution (8) describes a transient superradiant burst, with 
the maximal intensity at the delay time $t_0$, when
$y(t_0)=(z_0-1/g)^2(1+2\gm_3t_c)$, $z(t_0)=1/g$.
After this, for $t\gg t_0$, the coherence intensity exponentially 
diminishes and the spin polarization becomes inverted,
\be
\label{10}
 y\simeq 4y(t_0) e^{-2\gm_p t} \; , \qquad  z \simeq - z_0 + 2/g \; .
\ee
For sufficiently large coupling parameter $g$, the reversal of spin 
polarization is practically complete.

Now imagine that at some time after $t_0+\tau_p$ we again invert 
the spin polarization from that in Eq. (10) to the symmetric positive 
value. For large $g$, this inversion is practically from $-z_0$ to
$z_0$. Such an inversion can be realized in three possible ways:
inverting the external magnetic field $B_0$, acting on spins by a
resonant $\pi$-pulse, or just turning the sample $180^0$ about an
axis perpendicular to $B_0$. As a result, we get again a strongly 
nonequilibrium state of inverted spins. After the time $t_0$, counted
from the moment when the newly nonequilibrium state is prepared, 
another superradiant burst will arise. After the second burst dies out, 
one can again invert the spin polarization by one of the mentioned
methods. Then one more superradiant burst will appear. This procedure
can be repeated as many times as necessary for creating a required 
number of sharp superradiant pulses. The time intervals between bursts 
can be regulated, allowing the formation of different groups of pulses,
with varying intervals between separate groups. Thus, it is feasible to
compose a code, similar to the Morse alphabet, which can be used in 
processing information. It is this possibility of regulating temporal 
intervals between superradiant bursts which permits us to call the 
described phenomenon {\it punctuated spin superradiance}.

Spin superradiance can be realized in different materials under various 
experimental setups. Thus, it was observed on proton spins in propanediol
C$_3$H$_8$O$_2$, butanol C$_4$H$_9$OH, and ammonia NH$_3$ [5--8] and on
$^{27}$Al nuclear spins in ruby Al$_2$O$_3$ [9--11]. The characteristic
parameters for these experiments with nuclear spins are: the density
of spins $\rho\sim 10^{22}-10^{23}$ cm$^{-3}$, the Zeeman frequency
$\om_0\sim 10^8$ Hz, the spin-lattice relaxation $\gm_1\sim 10^{-5}$ 
s$^{-1}$, the spin-spin dephasing parameter $\gm_2\sim 10^5$ s$^{-1}$,
the dynamic broadening width $\gm_3\sim 10^4-10^5$ s$^{-1}$, the 
resonator ringing time $\tau\sim 10^{-6}$ s, that is, the resonator
damping $\gm\sim 10^6$ s$^{-1}$. For these values, the spin-resonator
coupling parameter $g$ in Eq. (6) varies between 10 and 100. As has 
been shown [21,22], if nuclear spins are inside a ferromagnet or
ferrimagnet, possessing long-range magnetic order, then the coupling
parameter $g$ can be increased by a factor of $\mu_B/\mu_N\sim 10^3$, 
where $\mu_B$ is the Bohr magneton and $\mu_N$ is the nuclear magneton.
Therefore, the coupling parameter $g$ can be made as large as 
$g\sim 10^5$. Among other materials, where spin superradiance could,
in principal, be observed, are granular magnets [23] and molecular
magnets [24]. In this way, there exists a large variety of materials
with different characteristics allowing for the optimal choice of 
parameters for realizing punctuated spin superradiance. Note that the
intervals between superradiant pulses can be varied in a very wide
diapason of the order of $T_1=\gm_1^{-1}$. For nuclear magnets, with
$\gm_1\sim 10^{-5}$ s$^{-1}$, this time $T_1$ may be as much as several
days. And for molecular magnets at low temperatures, it could range up 
to months.

To illustrate the feasibility of creating different groups of 
superradiant pulses, with varying time intervals, we solved Eqs. (7) 
numerically. Different time intervals are obtained by changing the
moments of polarization inversion. For the characteristic parameters
those are taken that are typical of nuclear magnets [5--11]. In 
particular, we take $\gm_1=10^{-5}$ s$^{-1}$, $\gm_2=10^5$ s$^{-1}$,
and $\gm_3$ is varied between $10^4$ and $10^5$ s$^{-1}$. The variation 
of $\gm_3$ even in a wider diapason does not essentially change the
picture. The coupling parameter $g$ has also been varied between $10$ 
and $10^3$; the whole picture being qualitatively the same, with the
main difference that for larger $g$ the spin inversion, according to
Eq. (10), is better. For the presentation in Figures 1 to 3, we set 
$g=10^3$. In the absence of dynamic nuclear polarization, $\sgm=-1$.
The resonator attenuation $\gm=10^6$ s$^{-1}$. Finally, as initial
conditions we take $y(0)=0$ and $z(0)=1$. The first of the latter tells
that at the initial time the transverse coherence is absent. That is, 
we consider a purely self-organized process when radiation coherence
arizes in a spontaneous manner. Figures 1 to 3 give some examples of
how it is possible to create different bunches of superradiant bursts.
The shown function $y(t)$ is proportional to radiation intensity. 
The meaning of this function, according to its definition (3), is to 
describe the level of coherence in the system. As is seen from the 
Figures, the maxima of superradiant bursts display a high level of 
coherence, almost reaching $100\%$. The time variable in the Figures 
is measured in units of $T_2=\gm_2^{-1}$. The short time scale is 
chosen here just for the convenience of presentation. As is explained
above, the same picture can be stretched to the time scale characterized
by $T_1$, which, for nuclear magnets, would range up to several days.
The first superradiant burst occurs at the delay time $t_0$. To simplify
figure captions, we accept the notation for the time intervals between 
the pulses of the $i$-group as $\tau_i$ and for the intervals between 
the $i$- and $j$- groups as $\tau_{ij}$. Figure 3 demonstrates that
a regime of equidistant superradiant pulses can also be realized. Such
a regime can be used for producing spin masers [13,25] operating in
pulsing superradiant mode.

In conclusion, we have demonstrated, both analytically and numerically, 
the feasibility of realizing the regime of punctuated spin superradiance. 
In this regime, one may form various groups of superradiant bursts, with 
different spacing between the pulses inside each group as well as with
different time intervals between the groups. The possibility of so 
punctuating spin superradiance can be employed for processing 
information.

This work was started when one of the authors (V.I.Y.) was visiting
the Ames Laboratory at the Iowa State University. He is very grateful 
to B.N. Harmon for his kind hospitality and many useful discussions. 
He also appreciates very much fruitful discussions and advice from 
B.C. Gerstein and M. Pruski. Financial support from the Iowa State 
University, USA, and from the S\~ao Paulo State Research Foundation, 
Brazil, is acknowledged.

\newpage

\begin{center}
{\large{\bf Figure Captions}}
\end{center}

{\bf Fig. 1}. Punctuated spin superradiance with three different groups
of superradiant bursts. All parameters are explained in the text. Time 
is measured in dimensionless units. The time intervals are: $t_0=0.029$,
$\tau_1=0.065$, $\tau_{12}=0.26$, $\tau_2=0.1$, $\tau_{23}=0.15$,   
$\tau_3=0.025$.

\vskip 5mm

{\bf Fig. 2}. Punctuated spin superradiance with four groups of 
superradiant pulses, the first group containing a single burst, the 
time intervals being: $t_0=0.024$, $\tau_{12}=0.226$, $\tau_2=0.025$, 
$\tau_{23}=0.35$, $\tau_3=0.075$, $\tau_{34}=0.275$, $\tau_4=0.1$.

\vskip 5mm

{\bf Fig. 3}. Punctuated spin superradiance with a comb of equidistant 
pulses, starting at $t_0=0.024$, having time intervals $\tau_1=0.1$.

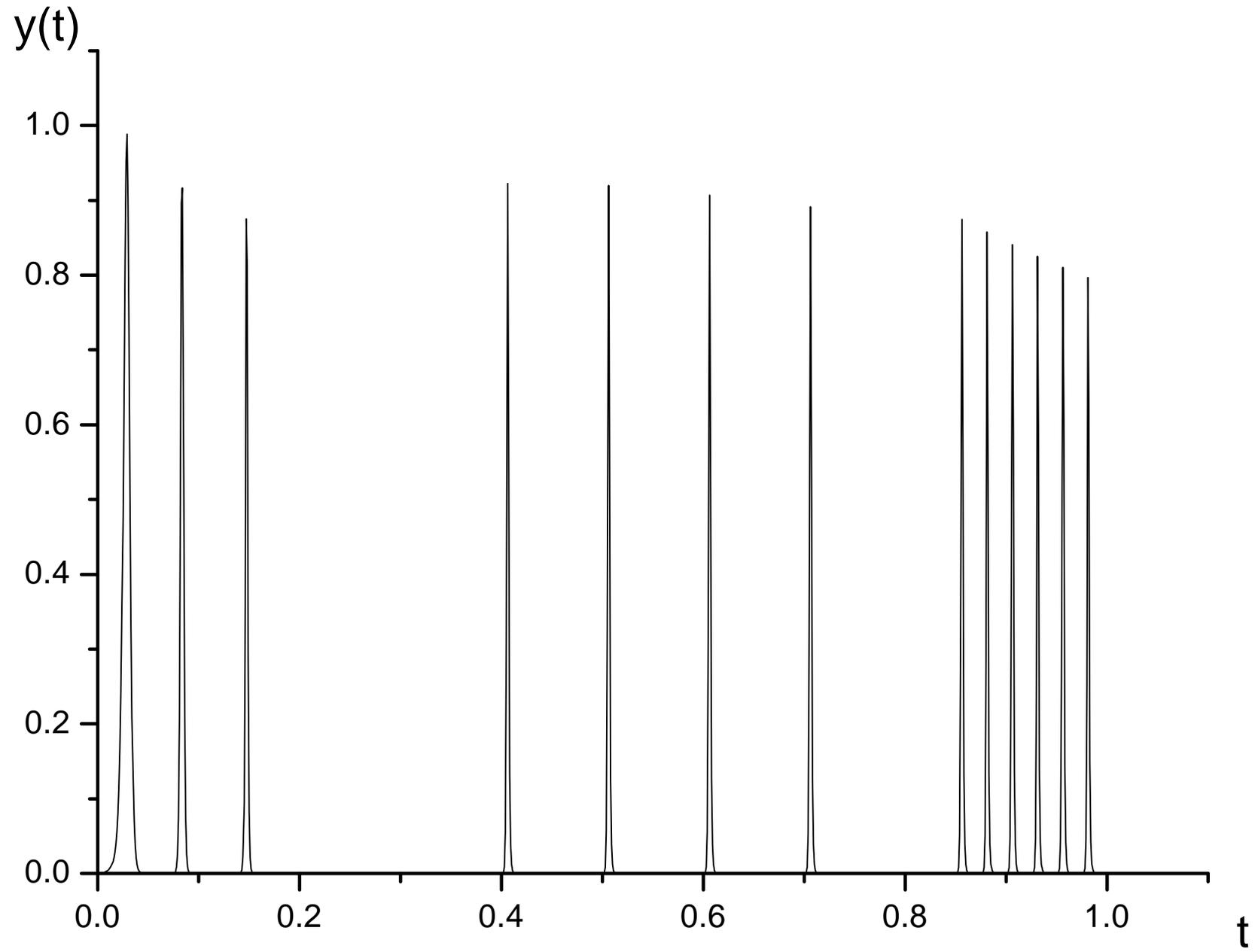
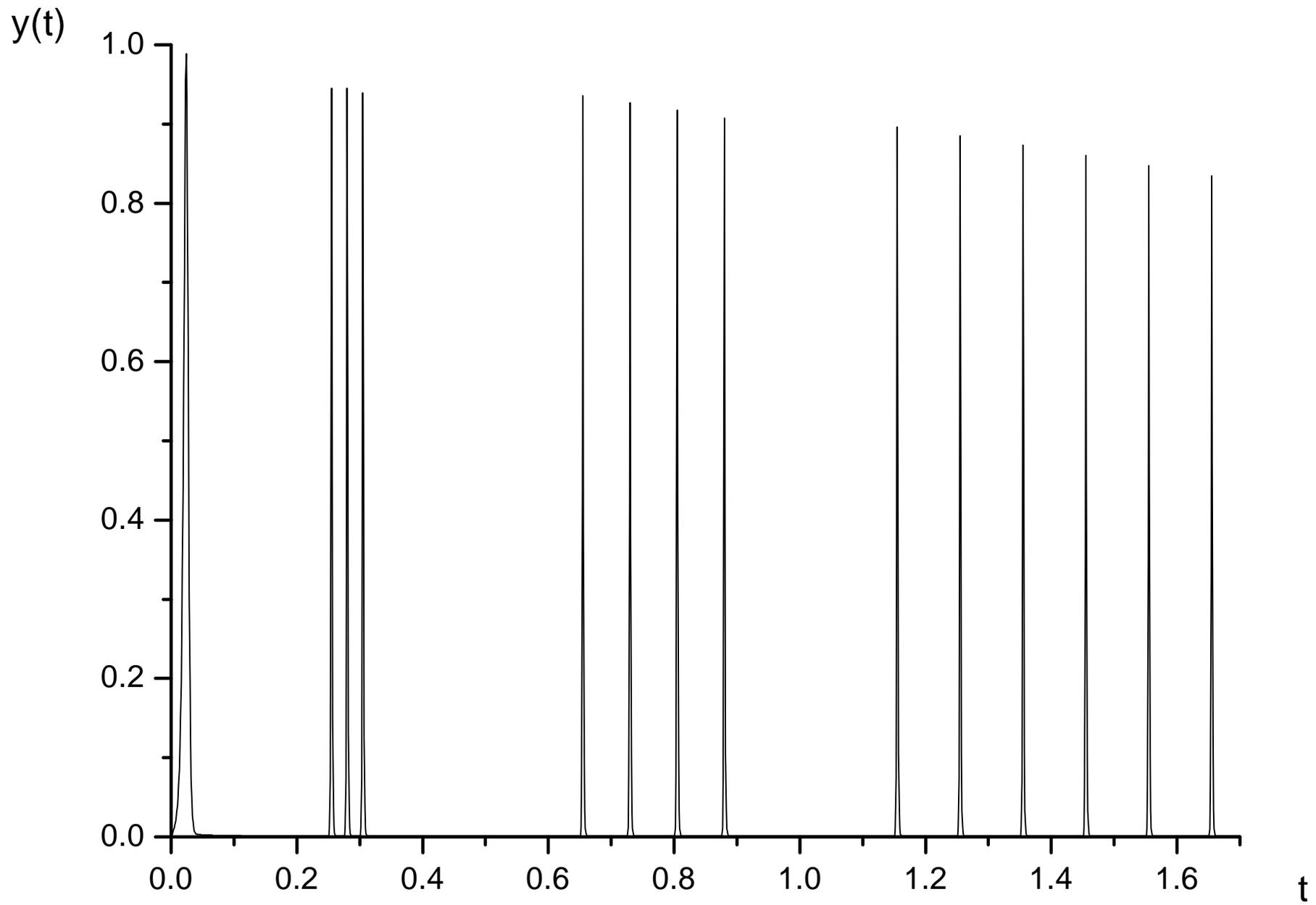
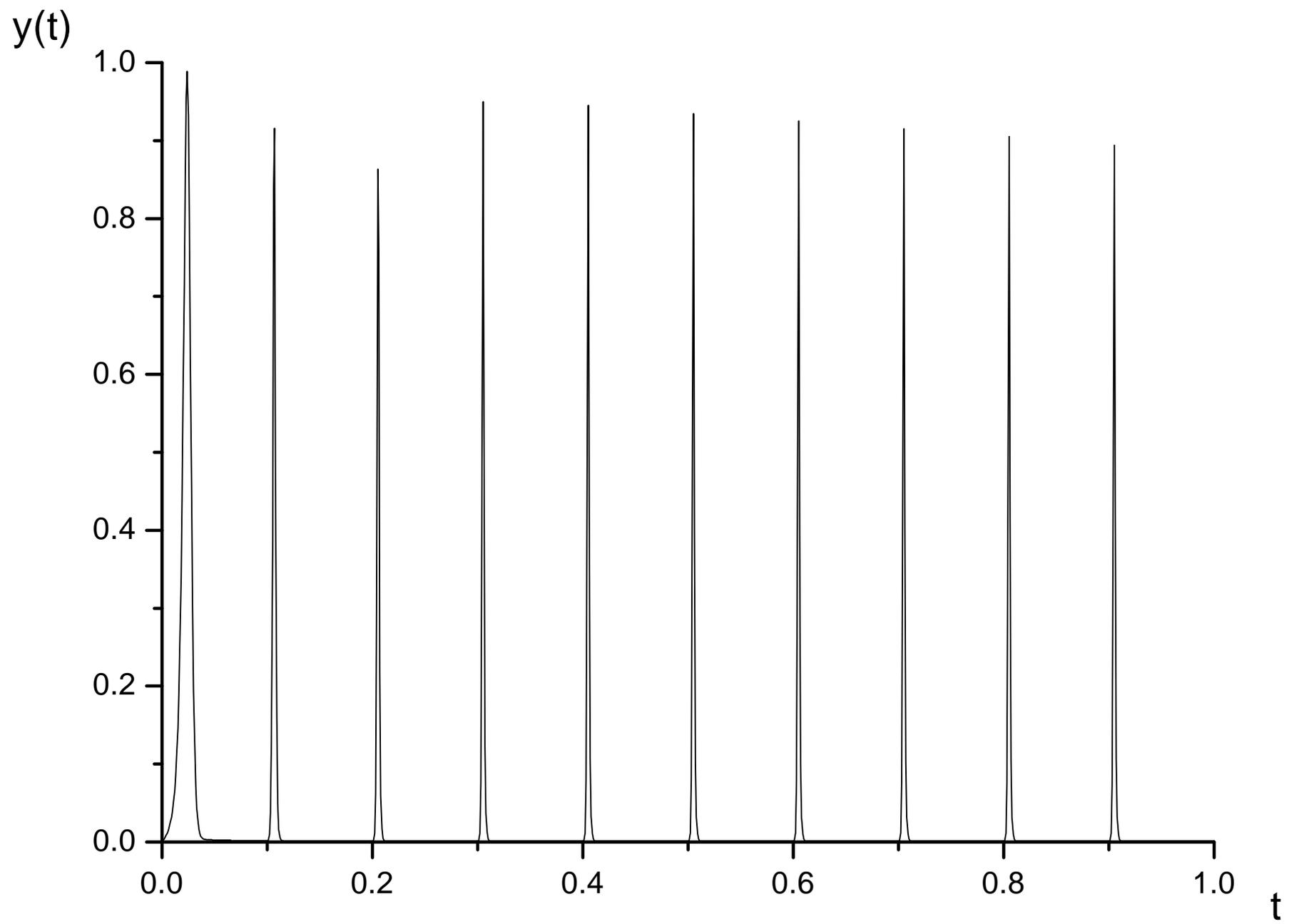
\end{document}